\documentclass[aps,prl,nofootinbib,longbibliography,superscriptaddress,twocolumn]{revtex4-2}
\usepackage{amsmath}
\usepackage{amssymb}
\usepackage{graphicx}
\usepackage{dcolumn}
\usepackage{bm}
\usepackage[colorlinks=true,linkcolor=blue,anchorcolor=blue, citecolor=blue,urlcolor=blue]{hyperref}
\usepackage[mathlines]{lineno}
\usepackage{ulem}
\usepackage{siunitx}
\usepackage{gensymb}

\makeatother

\begin{document}
	
\title{Pressure-Tunable Generalized Wigner Crystal and Fractional Chern Insulator in twisted MoTe$_2$}
\author{Bingbing Wang}
\thanks{These authors contributed equally to this work.}
\affiliation{Centre for Quantum Physics, Key Laboratory of Advanced Optoelectronic Quantum Architecture and Measurement (MOE), School of Physics, Beijing Institute of Technology, Beijing, 100081, China}
\affiliation{Beijing Key Laboratory of Quantum Matter State Control and Ultra-Precision Measurement Technology, Beijing 100081, China}
	
\author{Junxi Yu}
\thanks{These authors contributed equally to this work.}
\affiliation{Centre for Quantum Physics, Key Laboratory of Advanced Optoelectronic Quantum Architecture and Measurement (MOE), School of Physics, Beijing Institute of Technology, Beijing, 100081, China}
\affiliation{Beijing Key Laboratory of Quantum Matter State Control and Ultra-Precision Measurement Technology, Beijing 100081, China}
	
\author{Cheng-Cheng Liu}
\email{ccliu@bit.edu.cn}
\affiliation{Centre for Quantum Physics, Key Laboratory of Advanced Optoelectronic Quantum Architecture and Measurement (MOE), School of Physics, Beijing Institute of Technology, Beijing, 100081, China}
\affiliation{Beijing Key Laboratory of Quantum Matter State Control and Ultra-Precision Measurement Technology, Beijing 100081, China}

\begin{abstract}
	Due to the forming of low-energy flat bands, the moir\'e superlattices of the transition metal dichalcogenides are fascinating platforms for studying novel correlated states when such flat bands are fractionally filled, with the Coulomb interaction dominating. Here, we demonstrate that pressure can efficiently tune the flatness and quantum geometry of the single-particle bands in twisted bilayer MoTe$_2$ (\textit{t}MoTe$_2$). By fractionally filling the topmost valence band, we find that pressure can act as a flexible means to modulate the fractional Chern insulator (FCI) and the generalized Wigner crystal (GWC) and control their many-body topological phase transitions. Moreover, our results indicate a remarkable correspondence between the single-particle band geometry and the formation of FCI and GWC. As the recent experiments report the presence of FCI phases in \textit{t}MoTe$_2$, our predictions could be readily implemented experimentally.
\end{abstract}

\maketitle
\textit{Introduction.}---The moir\'e systems have attracted significant interest due to their low-energy flat bands with tunable multiple degrees of freedom, which enables extensive studies of topological and correlation effects~\cite{andreiGrapheneBilayersTwist2020,balentsSuperconductivityStrongCorrelations2020,cao_correlated_2018,cao_unconventional_2018,crepelAnomalousHallMetal2023,liSpontaneousFractionalChern2021,wangFractionalChernInsulator2024,dongTheoryQuantumAnomalous2024,guoFractionalChernInsulator2024,zhouFractionalQuantumAnomalous2024a,dongAnomalousHallCrystals2024,yvesh.kwanMoireFractionalChern,huangCorrelatedInsulatingStates2021,reganMottGeneralizedWigner2020,liImagingTwodimensionalGeneralized2021,xuCorrelatedInsulatingStates2020,sharmaTopologicalQuantumPhase2024,reddyFractionalQuantumAnomalous2023,crepelAnomalousHallMetal2023}. Recent experiments have reported the presence of fractional quantum anomalous Hall effect (FQAH), i.e., fractional Chern insulator (FCI), in moir\'e systems including twisted bilayer MoTe$_2$ (\textit{t}MoTe$_2$) and aligned pentalayer rhombohedral stacked graphene on hBN~\cite{zengThermodynamicEvidenceFractional2023,caiSignaturesFractionalQuantum2023,parkObservationFractionallyQuantized2023,luFractionalQuantumAnomalous2024,xuObservationIntegerFractional2023}. In the absence of the external magnetic field, strong Coulomb interaction leads to FCI with degenerate ground states separated by a many-body gap from the excited states~\cite{regnaultFractionalChernInsulator2011,zengThermodynamicEvidenceFractional2023,caiSignaturesFractionalQuantum2023,parkObservationFractionallyQuantized2023,luFractionalQuantumAnomalous2024,xuObservationIntegerFractional2023}.
On the other hand, strong repulsive interaction can also lead to the spontaneous breaking of translation symmetry, forming the Wigner crystal (WC)~\cite{wignerInteractionElectronsMetals1934}.

At low electron densities, electrons of the WC phase form a periodically ordered electron lattice.
In the 1970s, Hubbard defined the generalized Wigner crystal (GWC) that can localize electrons with the aid of atom lattice~\cite{hubbardGeneralizedWignerLattices1978}. GWC has been discovered in WSe$_2$/WS$_2$ moir\'e heterostructure~\cite{reganMottGeneralizedWigner2020,liImagingTwodimensionalGeneralized2021}. The WC was observed to compete with the FQH phase in Bernal stacked bilayer graphene with strong magnetic field~\cite{tsuiDirectObservationMagneticfieldinduced2024}, however, the coexistence of GWC and FCI within a single system remains elusive, and their fundamental interplay has yet to be systematically investigated. 

Investigating the competition mechanism between the two correlated phases necessitates comprehensive parameter space exploration. However, most conventional tuning approaches face fundamental limitations: either they have been thoroughly characterized (e.g., displacement fields) or require complete sample re-fabrication (e.g., twist angle adjustment).
Pressure, as a technically mature tuning approach~\cite{yankowitzTuningSuperconductivityTwisted2019,gaoBandEngineeringLargeTwistAngle2020a,szentpeteriTailoringBandStructure2021,zhangPressureTunableVan2022}, is especially powerful in engineering the moir\'e superlattice electronic properties. By tuning pressure, there is no need to fabricate different samples with different twist angles, which could avoid introducing disorder and reduce complexity and costs in device fabrication. Pressure was proven to be able to enhance the many-body gap of FQHE experimentally~\cite{morawiczEnhancement431993,morawiczFractionalQuantumHall1990,morawiczObservationFractionalQuantum1990}, and has been predicted to stabilize the FCI in twisted bilayer WSe$_2$~\cite{morales-duranPressureenhancedFractionalChern2023}.

In this work, we propose that pressure can act as a clean and efficient in situ knob to regulate the many-body topological phase transition between the GWC and FCI in \textit{t}MoTe$_2$. We use a pressure-parameterized continuous model and define the figure of merit to measure the flatness of the $n$-th energy band. We analyze the Berry curvature and quantum metric, which serve as geometric indicators for the emergence of FCI and GWC. Projected exact diagonalization (ED) results show the FCI and GWC characteristics within a certain parameter range at one-third filling ($\nu=\frac{1}{3}$). We further calculate the many-body Chern number, particle entanglement spectrum (PES), and spectral flows to determine the topological properties under different parameters. The static structure factor $S(\boldsymbol{q})$ is calculated to confirm the formation of GWC and pressure-tunable phase transition between GWC and FCI.

\textit{Continuum model under pressure.}---When the pressure applied along the vertical direction is not large enough to alter the symmetry of the system, the Hamiltonian for \textit{t}MoTe$_2$ reads~\cite{wuTopologicalInsulatorsTwisted2019}
\begin{equation}\label{eq:ContHam}
	H_0 (\boldsymbol{r})=\begin{pmatrix} \frac{\hbar^2\nabla^2}{2m^*}+V_{t}(\boldsymbol{r}) & t_\sigma(\boldsymbol{r}) \\ t_\sigma^{\dagger}(\boldsymbol{r}) & \frac{\hbar^2\nabla^2}{2m^*}+V_{b}(\boldsymbol{r}) \end{pmatrix}.
\end{equation}
$\sigma=\pm1$ is the spin index, which is locked to the valley degree of freedom, and $l$ is the layer index. The moir\'e potential $V_{\sigma,l}(\boldsymbol{r})$ and interlayer coupling $t_\sigma(\boldsymbol{r})$ are given by $V_{l}(\boldsymbol{r})=2V[\cos(\boldsymbol{G}_{1}^{M}\cdot \boldsymbol{r}+\psi_{l}) + \cos(\boldsymbol{G}_{2}^{M}\cdot \boldsymbol{r}+\psi_{l}) + \cos((\boldsymbol{G}_{1}^{M}+\boldsymbol{G}_{2}^{M})\cdot \boldsymbol{r}-\psi_{l})]$, and $t_\sigma(\boldsymbol{r})=\omega(1 + e^{-i\sigma \boldsymbol{G}_{2}^{M}\cdot \boldsymbol{r}} + e^{-i\sigma (\boldsymbol{G}_{1}^{M}+\boldsymbol{G}_{2}^{M})\cdot \boldsymbol{r}})$. $\boldsymbol{G}_{1/2}^{M}$ are the moir\'e reciprocal basic vectors. $m^*, V, \omega$ are the pressure-dependent parameters, whose specific functional forms can be derived from the local stacking approximation~\cite{anfaEffectiveHamiltonianTwisted2024}, as obtained through DFT calculations (see Supplementary Material (SM)~\cite{supplemental} for details).
Based on the continuum model above, we can modulate the bandwidth, band gap, band geometry, and Chern number of the low-energy bands of \textit{t}MoTe$_2$ as a function of pressure and twist angle. The physical origin of the evolution in band properties can be partially understood as changes in the moir\'e potential induced by the piezoelectric effect and ferroelectric polarization~\cite{zhangPolarizationdrivenBandTopology2024}. In addition, however, the changes in the electron effective mass also contributes (in the SM, we show the evolution of the Chern number as a function of the effective mass~\cite{supplemental}).

\begin{figure}[t]
	\includegraphics[width=0.5\textwidth]{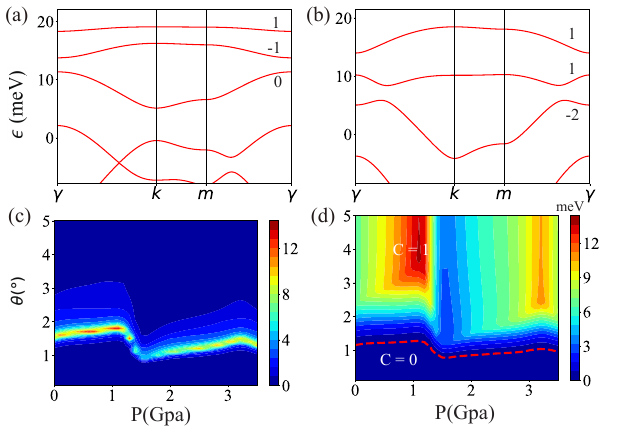}
	\caption{Band structures of \textit{t}MoTe$_2$ at (a) P=0 GPa; (b) P=1.8 GPa with twist angle 1.8\degree. (c) The figure of merit of the topmost valence band as a function of pressure and twist angle. The figure of merit reaches its maximum at approximately 1.8\degree and 0-1.0 GPa. (d) The topological phase diagram of the first valence band. Different colors represent the size of the local gap between the first and second bands. }\label{Band-qua}
\end{figure}

The band structures of P=0 GPa and 1.8 GPa are shown in Figs. \ref{Band-qua}(a) and (b) with twist angle 1.8\degree. Compared to P = 0 GPa, at P = 1.8 GPa, the gap between the first and second valence bands increases, while the first band becomes less flat. To better characterize bands' flatness, we define the corresponding figure of merit as follows
\begin{equation}
	\zeta_n\equiv\frac{\min(\Delta_{n-1,n},\Delta_{n,n+1})}{W_n},
\end{equation}
where $W_n$ is the bandwidth of the $n$-th band and $\Delta_{n,n+1}$ represents the global gap between the $n$-th and ($n+1$)-th bands. We use $\zeta_n$ to quantify the flatness of the bands, with bigger values indicating bands closer to the ideal flat-band limit. As shown in Fig. \ref{Band-qua}(c), the figure of merit of the first band takes a big value between 1\degree and 2\degree, where this band carries a nonzero Chern number and likely hosts the FCI state. The ratio of the global gap to the bandwidth reaches its maximum at around 1.8\degree. A key condition for realizing the FCI is that the band geometry closely resembles that of the lowest Landau level (LLL). We next analyze how pressure affects the band properties from the perspective of quantum geometry.

\textit{Band geometry.}---The geometric information is encapsulated in the quantum geometric tensor (QGT)~\cite{wangExactLandauLevel2021,ledwithVortexabilityUnifyingCriterion2023,royBandGeometryFractional2014}, which is defined as $\mathcal{Q_{\mu\nu}}(\boldsymbol{k})=\sum_{n\in occ}\langle\partial_{\mu}u_{n}(\boldsymbol{k})|1-P(\boldsymbol{k})|\partial_{\nu}u_{n}(\boldsymbol{k})\rangle \label{eq(1)}$, where $n$ is the index of occupied bands, and $P(\boldsymbol{k})\equiv\sum_{n\in occ}|u_{n}(\boldsymbol{k})\rangle\langle u_{n}(\boldsymbol{k})|$ is the projection operator. The QGT can be further decomposed into the real and imaginary parts $\mathcal{Q_{\mu\nu}}(\boldsymbol{k})=g_{\mu\nu}(\boldsymbol{k})-\frac{i}{2}\Omega_{\mu\nu}(\boldsymbol{k})$, where $g_{\mu\nu}(\boldsymbol{k})$ is quantum metric and $\Omega_{\mu\nu}(\boldsymbol{k})$ is Berry curvature.  For the $n$-th Landau level (LL),  $\Omega_{xy}(\boldsymbol{k})=l_{B}^{2}$, $g_{xx}(\boldsymbol{k})=g_{yy}(\boldsymbol{k})=(n+1/2)l_{B}^{2}$, $g_{xy}(\boldsymbol{k})=g_{yx}(\boldsymbol{k})=0$~\cite{ozawaRelationsTopologyQuantum2021,ledwithStrongCouplingTheory2021,peottaSuperfluidityTopologicallyNontrivial2015}, where $l_{B}=\sqrt{\hbar c/eB}$ refers to the magnetic length. As one can see, both $\Omega$ and $g$ are constants in the $n$-th LL. For LLL and any two-level system, $\Omega_{xy}$ and $g$ satisfy the relation $\mathrm{tr}(g)=|\Omega_{xy}|$,
which is called the trace condition~\cite{ledwithFractionalChernInsulator2020,ledwithStrongCouplingTheory2021}. Therefore, to effectively simulate the LLL, the $\Omega_{xy}$ and tr$(g)$ of the topmost \textit{t}MoTe$_2$ valence band are as uniform as possible. Their fluctuations over the moir\'e Brillouin zone (mBZ) can be quantified by the standard deviations,
$
\sigma(\Omega_{xy})\equiv\sqrt{\langle\Omega_{xy}^{2}\rangle-\langle\Omega_{xy}\rangle^{2}},
\sigma(\mathrm{tr}(g))\equiv\sqrt{\langle\mathrm{tr}(g)^{2}\rangle-\langle\mathrm{tr}(g)\rangle^{2}},
$ where $\langle\mathcal{O}\rangle$ is the average value of the operator $\mathcal{O}$ over the mBZ.
Furthermore, we can also measure the deviation of the band geometry from the trace condition, denoted as $T=(1/2\pi)\int_{mBZ}d^{2}\boldsymbol{k}\ (tr(g(\boldsymbol{k}))-|\Omega_{xy}(\boldsymbol{k})|)$~\cite{ledwithFractionalChernInsulator2020}.

\begin{figure}[t]
	\includegraphics[width=0.5\textwidth]{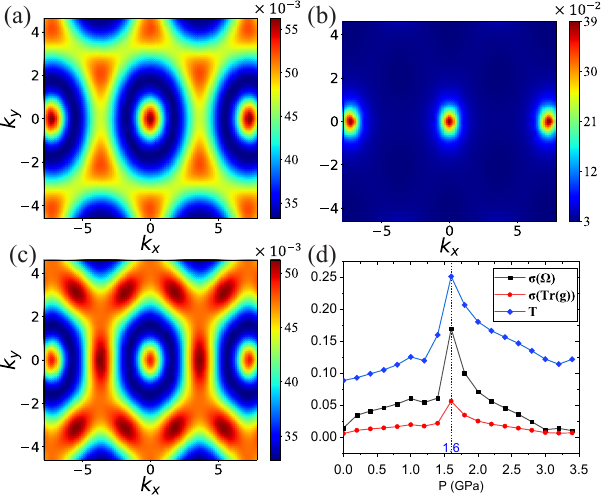}
	\caption{Band geometry of the topmost band with the 1.8\degree twist angle as a function of pressure. (a)-(c) The Berry curvature distribution at (a) P=0 GPa; (b) P=1.8 GPa; (c) P=3.0 GPa. (d) Fluctuations of Berry curvature $\sigma(\Omega_{xy})$, quantum metric trace $\sigma(\mathrm{tr}(g))$, and deviation of the band geometry from trace condition $T$ as functions of pressure.}
	\label{Geo}
\end{figure}

\begin{figure*}[t]
	\includegraphics[width=1.0\textwidth]{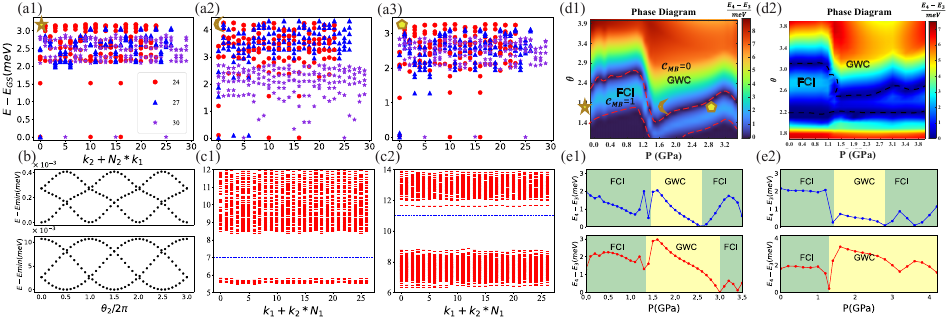}
	\caption{Many-body low energy spectra at $\nu=\frac{1}{3}$ for (a1) P=0.0 GPa; (a2) P=1.8 GPa; (a3) P=3.0 GPa with twist angle 1.8\degree and $\epsilon=10$. The red dots,  blue triangles, and blue-violet stars label the spectra of the clusters of 24, 27, and 30 sites, respectively. The parameters in (a1), (a2), and (a3) correspond to the points marked by star, crescent, and pentagon symbols in the phase diagram (d1), respectively. (b) Spectral flow of the 30-site cluster at P=0 GPa (top panel)and P=3.0 GPa (bottom panel). The flux insertion is along the direction of $\boldsymbol{T_2}$. (c1)(c2) PES for the 27-site with $N_{A}$ = 3 at 1.8 GPa (c1) and 3.0 GPa (c2). The numbers of PES levels below the dashed line are 252 and 1710, respectively. (d1) The phase diagram determined by the many-body gap $E_4-E_3$ as a function of $\theta$ and P, calculated on the 27-site. $\mathcal{C}_{MB}$ labels the many-body Chern number. (e1) The many-body gap versus pressure along line cuts in (d1) at twist angle 1.8\degree (top panel) and 2.0\degree (bottom panel). (d2) Phase diagram considering lattice relaxation effect. (e2) The many-body gap versus pressure along line cuts in (d2) at twist angle 2.6\degree (top panel) and 2.9\degree (bottom panel).}
	\label{ED 1/3}
\end{figure*}

We select the 1.8\degree ~twist angle, since the figure of merit is remarkably high between 1.0 GPa-1.3 GPa as plotted in Fig. \ref{Band-qua}(c). As shown in Figs. \ref{Geo}(a)(c), at P = 0 GPa and 3.0 GPa, the Berry curvature $\Omega_{xy}$ is distributed uniformly across the mBZ. In contrast, at P = 1.8 GPa, this uniformity is disrupted, with $\Omega_{xy}$ peaking at the $\gamma$ point (Fig. \ref{Geo}(b)). The distribution of the quantum metric trace tr$(g)$ presents a similar behavior to $\Omega_{xy}$, which is plotted in SM~\cite{supplemental}. As shown in Fig. \ref{Geo}(d), $\sigma(\Omega_{xy})$, $\sigma(tr(g))$, and $T$ all peak around middle pressure (1.6 GPa) and exhibit small values at the two sides. These results suggest that the band geometry at low and high pressures better resembles the LLL, favoring the FCI, while intermediate pressures promote competing phases.

\textit{Many-body spectrum.}---The ED method is employed to investigate the FCI and GWC phases. In this work, we consider the two-body long-range Coulomb interaction $V_{C}(\boldsymbol{r})=\frac{e^{2}}{\epsilon|r|}$, whose Fourier transformed form is $V_{C}(\boldsymbol{q})=\frac{2\pi e^{2}}{\epsilon|\boldsymbol{q}|}$. Projecting the Coulomb interaction to the topmost band of \textit{t}MoTe$_2$, the total Hamiltonian can be expressed as $\tilde{H}=H_0+H_{int}$,  among which
\begin{equation}
	H_{int}=\frac{1}{2A}\sum_{\{\boldsymbol{k_{i}}\},\sigma,\sigma'}\sum_{\boldsymbol{q}}V^{{\boldsymbol{k_{1}k_{3}},\sigma}}_{\boldsymbol{k_{2}k_{4}},\sigma'}(\boldsymbol{q})\gamma_{\boldsymbol{k_{1}},\sigma}^{\dag}\gamma_{\boldsymbol{k_{2}},\sigma'}^{\dag}\gamma_{\boldsymbol{k_{4}},\sigma'}\gamma_{\boldsymbol{k_{3}},\sigma},
\end{equation}
and $H_0=-\sum_{\boldsymbol{k}\in mBZ,\sigma}\epsilon_{\boldsymbol{k},\sigma}\gamma_{\boldsymbol{k},\sigma}^{\dag}\gamma_{\boldsymbol{k},\sigma}$.
The minus sign in front of the $\epsilon_{k,\sigma}$ term is due to our choice of hole filling, and $A$ is the cluster area. 
$\gamma_{\boldsymbol{k},\sigma}^{\dag}=\sum_{\boldsymbol{G},l}u_{\boldsymbol{G},l,\sigma}(\boldsymbol{k})c_{\boldsymbol{k}+\boldsymbol{G},l,\sigma}^{\dag}$ represents the creation of a hole in the topmost valence band. The specific form of $V^{{\boldsymbol{k_{1}k_{3}},\sigma}}_{\boldsymbol{k_{2}k_{4}},\sigma'}(\boldsymbol{q})$ is provided in SM~\cite{supplemental}.

We first select the cluster cell of $N_{\Phi}=N_1\times N_2$ with three sizes $N_{\Phi_{1}}=4\times6=24$, $N_{\Phi_{2}}=9\times3=27$ and $N_{\Phi_{3}}=5\times6=30$. The reciprocal basis vectors are denoted as $\boldsymbol{T}_{1}$ and $\boldsymbol{T}_{2}$. The center of mass (COM) momentum of $N_e$ particles is given by $\boldsymbol{K}_\mathrm{COM}=k_{1}\boldsymbol{T}_{1}+k_{2}\boldsymbol{T}_{2}$, and can be denoted as $K_\mathrm{COM}=k_{1}+N_{1}k_{2}$ ($k_{i}=0,...,N_{i}-1$). The $K_\mathrm{COM}$ positions corresponding to each cluster shape are provided in SM~\cite{supplemental}.
Due to the translation symmetry of torus geometry, the FCI state will exhibit ground states with $q$-fold degeneracy when $\nu=\frac{p}{q}$, with coprime integers $p$ and $q$~\cite{regnaultFractionalChernInsulator2011,wenGroundstateDegeneracyFractional1990}. The GWC phase may exhibit the same degenerate ground states as the FCI in some cases, depending on how spatial translation symmetry is broken~\cite{wilhelmInterplayFractionalChern2021}. At twist angle 1.8\degree, we calculate the low-energy many-body ED spectrum assuming full spin polarization at $\nu=\frac{1}{3}$ as shown in Figs. \ref{ED 1/3}(a1)-(a3). For the clusters of $N_{\Phi_1}=24$ and $N_{\Phi_3}=30$, the three degenerate ground states located at momentum of $K_\mathrm{COM}$ = (0, 10, 16) and (5, 15, 25) at 0 GPa, respectively, which satisfies the generalized Pauli principle~\cite{haldaneFractionalStatisticsArbitrary1991a,supplemental}. As the pressure increases, the positions of degenerate states for the 24-site cluster remain unchanged. However, the case is different for $N_{\Phi_{2}}=27$, where $N_e$ is commensurate with the cluster dimensions, and the three topological states have the same momenta in the FCI phase~\cite{supplemental}. As shown by the blue triangles in Figs. \ref{ED 1/3}(a1)-(a3), the degeneracy locates at $\gamma$ at 0 GPa and 3.0 GPa, while it distributes to $\gamma$, $k_\pm$ at 1.8 GPa. This indicates a new competing phase, which we consider as GWC, as demonstrated in the following. In contrast to the 27-site and 24-site cases, there is no many-body gap at P = 1.8 Gpa for the 30-site case, indicating that the formation of GWC is closely related to the manner of spontaneous translation symmetry breaking. From the perspective of composite fermions (CF), the commensurate filling factors follow the Jain sequence $\nu=\frac{p}{2 pm+1}(m\in \mathbb{Z})$~\cite{jainCompositefermionApproachFractional1989}.
We also choose the filling of $\nu=\frac{2}{5}$ in cluster shape of $N_{\Phi_{4}}=5\times5$, which shows the similar results as $\nu=\frac{1}{3}$  (see the SM\cite{supplemental}).

\textit{Many-body topological properties.}---The topological properties of ground states can be characterized by the many-body Chern number under twist boundary conditions~\cite{niuQuantizedHallConductance1985,goldmanRelatingHallConductivity2024,shengFractionalQuantumHall2011}, which are implemented by superimposing a twisting phase angle on the original crystal lattice translation symmetry. This operation is given by $\hat{T}_{M}(\boldsymbol{L_{i}})|\psi(\boldsymbol{r_{a}})\rangle=e^{i\theta_{i}}|\psi(\boldsymbol{r_{a}})\rangle$, with $\hat{T}_{M}$ being the magnetic translation operator. The effect of the twisted phase angle $\theta_{i}$ can be averaged over each $\boldsymbol{k}$ point, which is equivalent to transform $\boldsymbol{k}\rightarrow\boldsymbol{k}+\boldsymbol{\delta{k}}$ with $\boldsymbol{\delta{k}}=\sum_{i}\frac{\theta_{i}}{2\pi}\boldsymbol{T_{i}}$. Applying a 2$\pi$ phase in the $i$-direction is equivalent to inserting a quantum flux into one of the tubes of the torus manifold.
\begin{table}[b]
	\caption{\label{Chernnumber}The many-body Chern numbers of the three degenerate ground states labeled as 1, 2, and 3 are calculated for the 24-site and 27-site (values inside the parentheses) clusters. The twist angle is 1.8\degree.}
	
	\begin{ruledtabular}
		\begin{tabular}{c| c c c}
			Pressure(GPa)&1&2&3\\
			\colrule
			0&-0.333(-1)&-0.333(0)&-0.333(0)\\
			1.6&0.02(0)&0.02(0)&-0.04(0)\\
			1.8&0.009(0)&0.009(0)&-0.019(0)\\
			3.0&-0.333(0)&-0.333(0)&-0.333(-1)\\
		\end{tabular}
	\end{ruledtabular}
\end{table}
The expression for the many-body Chern number is~\cite{niuQuantizedHallConductance1985,shengFractionalQuantumHall2011}
\begin{equation}
	\mathcal{C}_\mathrm{MB}=\frac{1}{2\pi}\int_{0}^{2\pi}d\theta_{1}\int_{0}^{2\pi}d\theta_{2}F(\theta_{1},\theta_{2}).
\end{equation}
$F(\theta_{1},\theta_{2})=\mathrm{Im}(\langle\frac{\partial\Psi(\theta_{1},\theta_{2})}{\partial\theta_{2}}|\frac{\partial\Psi(\theta_{1},\theta_{2})}{\partial\theta_{1}}\rangle-c.c.)$ represents the many-body Berry curvature, where $\Psi(\boldsymbol{\theta})$ is the many-body ground state and can be written as a linear combination of the many-body orbital basis, which is formed as a direct product of single-particle states. Thus, the many-body Chern number includes contributions from both the many-body and single particle wave functions~\cite{supplemental,shengQuantumAnomalousHall2024}.

The many-body Chern number $\mathcal{C}_\mathrm{MB}$ for the three degenerate states of the 24-site and 27-site clusters at different pressures are shown in TABLE \ref{Chernnumber}. At P = 0 GPa and 3.0 GPa, the average $\mathcal{C}_\mathrm{MB}$  of each degenerate state is approximately -1/3, while at P = 1.8 GPa, the total $\mathcal{C}_\mathrm{MB}$ is 0. Therefore, the FCI state emerges at P = 0 and 3.0 GPa. In the next section, we show that near P = 1.8 GPa the correlated insulating states with threefold groundstate degeneracy and zero many-body Chern number is GWC. The PES (Figs. \ref{ED 1/3}(c1)(c2)) results further support this conclusion. The total density matrix is defined as $\rho=\frac{1}{N_{GS}}\sum_{i}|GS_{i}\rangle\langle GS_{i}|$, where the $N_{GS}$ is the degeneracy. The number of low-lying PES levels below the gap matches the 1/3 FQHE quasihole excitation counting rule~\cite{regnaultFractionalChernInsulator2011,sterdyniakExtractingExcitationsModel2011a,chandranBulkedgeCorrespondenceEntanglement2011,hermannsHaldaneStatisticsFinitesize2011,xuMultipleChernBands2025} at P = 3.0 GPa, but violates the rule at 1.8 GPa (more data are included in SM~\cite{supplemental}). When the flux is applied in one direction, the ground states evolve into each other~\cite{thoulessLevelCrossingFractional1989}, and the spectral flow is maintained as shown in Fig. \ref{ED 1/3}(b). 
At $\theta_{i}=3\times2\pi$, the energy states return to their original configuration, and the average Hall conductivity of each state is $\sigma_{xy}=-\frac{1}{3}e^2/h$. We present the complete phase diagram in Fig. \ref{ED 1/3}(d1), and line cuts at 1.8\degree $\ $and 2\degree $\ $in Fig. \ref{ED 1/3}(e1). The results demonstrate that pressure not only tunes the phase transition between the FCI and GWC, but also enhances the many-body gap with some twist angles(e.g., around 2\degree).

\begin{figure}[t]
	\includegraphics[width=0.5\textwidth]{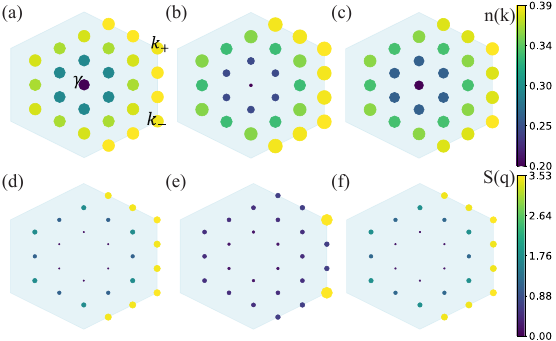}
	\caption{The occupation number $n(\boldsymbol{k})$ at (a) P=0 GPa with high symmetry points in the mBZ; (b) P=1.8 GPa; (c)P=3.0 GPa at 1.8\degree. The size and color of the circles together represent the magnitude of the value. (d)-(f) The structure factor $S(\boldsymbol{q})$ corresponding to the parameters in (a)-(c). }
	\label{S_n}
\end{figure}

\textit{Charge density order.}---The variation in the momentum-space distribution of degenerate states enhances the resolution of phase transition identification. Consequently, the 27-site cluster is utilized to examine the distinctive features of the GWC phase. To further verify the emergence of the GWC phase, the momentum space occupation $n(\boldsymbol{k})$ and static charge structure factor $S(\boldsymbol{q})$ are calculated. $n(\boldsymbol{k})$ represents the particle distribution in the mBZ, which can be expressed as
\begin{equation}
	n(\boldsymbol{k})=\frac{1}{N_{GS}}\sum_{i\in N_{GS}}\langle GS_{i}|\gamma_{\boldsymbol{k}}^{\dagger}\gamma_{\boldsymbol{k}}|GS_{i}\rangle,
\end{equation}
where $|GS_{i}\rangle$ denotes the nearly degenerate many-body ground states. At P = 0 GPa and 3 GPa, $n(\boldsymbol{k})$ (Figs. \ref{S_n}(a)(c)) and $\Omega_{xy}(\boldsymbol{k})$ (Figs. \ref{Geo}(a)(c)) are uniformly distributed in the mBZ, allowing the Berry curvature to effectively act on the holes, similar to the FQHE mechanism. However, in Fig. \ref{S_n}(b), at P = 1.8 GPa, the holes are pushed from the mBZ center to the boundary, while the Berry curvature remains concentrated at the $\gamma$ point (Fig. \ref{Geo}(b)), weakening this coupling. In this case, the holes form a spatially ordered state with an enlarged lattice constant to reduce the Coulomb repulsion.

The static structure factor $S(\boldsymbol{q})$, which can describe the charge-ordered structure and be measured in X-ray scattering experiments, reads
\begin{equation}
	S(\boldsymbol{q})=\frac{1}{N_{GS}}\sum_{i\in N_{GS}}\frac{1}{N_{\Phi}}\{\langle\rho(\boldsymbol{q})\rho(\boldsymbol{-q})\rangle-\langle\rho(\boldsymbol{q})\rangle^{2}\delta_{\boldsymbol{q,0}}\}.
\end{equation}
Here $\langle\rho(\boldsymbol{q})\rho(\boldsymbol{-q})\rangle$ denotes $\langle GS_{i}|\rho(\boldsymbol{q})\rho(\boldsymbol{-q})|GS_{i}\rangle$. As shown in Fig. \ref{S_n}(e), at 1.8 GPa, the peak values of $S(\boldsymbol{q})$ appear at $k_{\pm}$ in the mBZ. This indicates a lattice reconstruction caused by the ordered local structure of electrons in real space, which corresponds to the $\sqrt{3}\times\sqrt{3}$ GWC phase. We calculate the dimensionless density parameter defined in ref \cite{padhiGeneralizedWignerCrystallization2021} $r_{s}\approx$ 29.2, which is close to the critical threshold for Wigner crystallization ($r_{s}\approx$ 30) and larger than the experimentally observed value~\cite{liImagingTwodimensionalGeneralized2021}. Due to the moir\'e potential, the threshold could be lower than 30. However, the FCI phase (Figs. \ref{S_n}(d)(f)) shows a relatively uniform $S(\boldsymbol{q})$ distribution.

\textit{Discussion.}---In summary, our key finding is that the FCI and GWC phases in the \textit{t}MoTe$_2$ system are pressure-tunable in terms of band figure of merit, band geometry, many-body spectrum, spectral flow, PES, many-body Chern number, momentum space occupation number, and structure factor. Specifically, at certain twist angles, the \textit{t}MoTe$_2$ system transitions from the FCI phase to the GWC phase and subsequently reverts to the FCI phase, as pressure increases from 0 GPa to 3.0 GPa. Similar pressure of several GPa has been utilized to engineer the properties of graphene moir\'e superlattices~\cite{yankowitzTuningSuperconductivityTwisted2019,gaoBandEngineeringLargeTwistAngle2020a,szentpeteriTailoringBandStructure2021,zhangPressureTunableVan2022}. In the phase diagram in Fig. \ref{ED 1/3}(d1), the phase boundary between the FCI and GWC phases shows a general consistency with our defined figure of merit to describe the flatness of bands (Fig. \ref{Band-qua}(c)). This consistency confirms that band flatness, quantum geometry, and interactions collectively govern the emergence of the FCI and GWC, as well as their phase transitions. Pressure could exhibit a stabilizing effect on the FCI phase in \textit{t}MoTe$_2$.

 We further considering the effect of relaxation on \textit{t}MoTe$_2$. Our recent work~\cite{yu2025relaxationeffectselectronicstructure} derived the the relaxation effects in moir\'e materials through an analytical framework, effectively describing the impact of relaxation on electronic band structures. We briefly summarized the main conclusions in the SM~\cite{supplemental} and applied them to pressurized \textit{t}MoTe$_2$. The many-body phase diagram considering relaxation effect is shown in Fig.~\ref{ED 1/3}(d2). Compared with the scenario without relaxation, the FCI-GWC transition angle at ambient pressure increases to approximately 2.8\degree. At 2.4\degree, a reentrant phase transition of FCI-GWC-FCI can still be observed (Fig.~\ref{ED 1/3}(e2)). Exploring the impact of pressure on the non-Abelian effects in the fractionally filled second moir\'e valence band represents an exciting avenue for future research.

\begin{acknowledgments}
	\textit{Acknowledgments.}--- We thank Prakash Sharma for stimulating discussions. The work is supported by the NSF of China (Grant No. 12374055), the Science Fund for Creative Research Groups of  NSFC (Grant No. 12321004), and the National Key R\&D Program of China (Grant No. 2020YFA0308800).
\end{acknowledgments}

\bibliography{reference}

\clearpage
\onecolumngrid
\begin{center}
	\textbf{\large Supplementary Material for ``Pressure-Tunable Generalized Wigner Crystal and Fractional Chern Insulator in twisted MoTe$_2$"}\\[.2cm]
\end{center}

\maketitle
\setcounter{equation}{0}
\setcounter{figure}{0}
\setcounter{table}{0}
\setcounter{page}{1}
\renewcommand{\theequation}{S\arabic{equation}}

\renewcommand{\thefigure}{S\arabic{figure}}
\renewcommand{\thetable}{\arabic{table}}
\renewcommand{\tablename}{Supplementary Table}

\renewcommand{\bibnumfmt}[1]{[S#1]}
\renewcommand{\citenumfont}[1]{#1}
\makeatletter

\maketitle

\setcounter{equation}{0}
\setcounter{section}{0}
\setcounter{figure}{0}
\setcounter{table}{0}
\setcounter{page}{1}
\renewcommand{\theequation}{S-\arabic{equation}}

\renewcommand{\thefigure}{S\arabic{figure}}
\renewcommand{\thetable}{\arabic{table}}
\renewcommand{\tablename}{Supplementary Table}

\renewcommand{\bibnumfmt}[1]{[S#1]}
\makeatletter

\maketitle
\section{Continuum Model of tMoTe$_2$ under pressure}
Here, we introduce the details of the single-particle model for the twisted system. The Hamiltonian that we use to describe the low-energy physics of twisted TMDs takes the following form
\begin{equation}
	H_0(\mathbf{r})=
	\left(
	\begin{array}{cc}
		-\frac{\hbar^2\nabla^2}{2m^{*}}+V_{t}(\mathbf{r}) & t_{\sigma}(\mathbf{r})\\
		t_{\sigma}^{\dagger}(\mathbf{r}) & -\frac{\hbar^2\nabla^2}{2m^{*}}+V_{b}(\mathbf{r})
	\end{array}
	\right).
\end{equation}
$V_{l}(\mathbf{r})$ and $t_{\sigma}(\mathbf{r})$ represent the intralayer moir\'e potential and the interlayer coupling respectively, and they can be specifically written as
\begin{equation}
	\begin{aligned}
		V_{l}(\boldsymbol{r})&=2V[\cos(\boldsymbol{G}_{1}^{M}\cdot \boldsymbol{r}+\psi_{l}) + \cos(\boldsymbol{G}_{2}^{M}\cdot \boldsymbol{r}+\psi_{l}) \\
		&\quad + \cos((\boldsymbol{G}_{1}^{M}+\boldsymbol{G}_{2}^{M})\cdot \boldsymbol{r}-\psi_{l})],\\
		t_\sigma(\boldsymbol{r})&=\omega(1 + e^{-i\sigma \boldsymbol{G}_{2}^{M}\cdot \boldsymbol{r}} + e^{-i\sigma (\boldsymbol{G}_{1}^{M}+\boldsymbol{G}_{2}^{M})\cdot \boldsymbol{r}}).
	\end{aligned}
\end{equation}
$m^*$ represents the effective mass that describes the band dispersion near the $\mathbf{K}_{+}$ valley of a monolayer $\rm MoTe_2$. The index $l=\{t,b\}$ represents the layer index, where t and b denote the upper layer and the lower layer respectively. $\sigma=\pm$ is the valley index (since spin-valley locking exists, it is also the spin index). $\boldsymbol{G}^M_i=\frac{4\pi}{\sqrt{3}a_M}[\mathrm{cos}(\frac{2\pi(i-1)}{3}),\mathrm{sin}(\frac{2\pi(i-1)}{3})]$ represents the reciprocal lattice vectors of the twisted system. In this work, we choose $a_M=$ 3.52 \AA. The above Hamiltonian can be solved through plane-wave expansion.

Generally, under the local stacking approximation, the continuum model parameters for twisted systems can be obtained by calculating untwisted bilayer structures~\cite{PhysRevB.89.205414, PhysRevB.111.075434}. Specifically, for MoTe$_2$, the continuum model parameters are determined by performing DFT calculations on bilayer MoTe$_2$ with different interlayer displacements and fitting the band structure at the $\mathrm{K}_{\pm}$-points~\cite{PhysRevLett.122.086402}. Similarly, we can analyze and fit the band structures of shifted bilayer MoTe$_2$ under different vertical pressures to obtain pressure-dependent parameters. In this work, we directly adopt the results from Ref. \cite{anfaEffectiveHamiltonianTwisted2024}.

The moir\'e potential coefficient can be expressed as
\begin{equation}
	V(p) = [\frac{a+bp+cp^2+dp^3+ep^4+fp^5}{g+hp+ip^2+jp^3+kp^4+lp^5}] \rm meV,
\end{equation}
where
\begin{equation}
	\begin{aligned}
		a &= -694, & b&=258, & c&=1327, & d&=-1405, \\
		e &= 486.1, & f&=-55.8, & g&=-80.5, & h&=79.7, \\
		i &= 61.9, & j&=-102.3, & k&=39.51, & l&=-4.783.
	\end{aligned}
\end{equation}
The phase parameter in the moir\'e potential remains pressure-independent: $\psi(p)_{t/b}=\mp90.0797^{\circ}$. The interlayer coupling coefficient is given by
\begin{equation}
	\omega(p) = (-8.35-1.58p+0.25p^2) \rm{meV},
\end{equation}
and the effective mass can be written as
\begin{equation}\label{eq:effmass}
	m^{*}(p) = (0.64-0.193p+0.0335p^2)m_{e}.
\end{equation}
In all expressions above, the pressure 
$p$ is in units of $\rm Gpa$.
Additionally, recent studies have shown that in-plane relaxation shows discernible involvement in twisted bilayer $\rm MoTe_2$~\cite{reddyFractionalQuantumAnomalous2023,jiaMoireFractionalChern2024,wangFractionalChernInsulator2024}. However, this effect leads to the breakdown of the local stacking approximation, thereby substantially increasing the complexity of the investigation. As an exploratory study, we therefore do not consider in-plane relaxation in our current work.

\begin{figure*}[h]
	\includegraphics[width=0.5\textwidth]{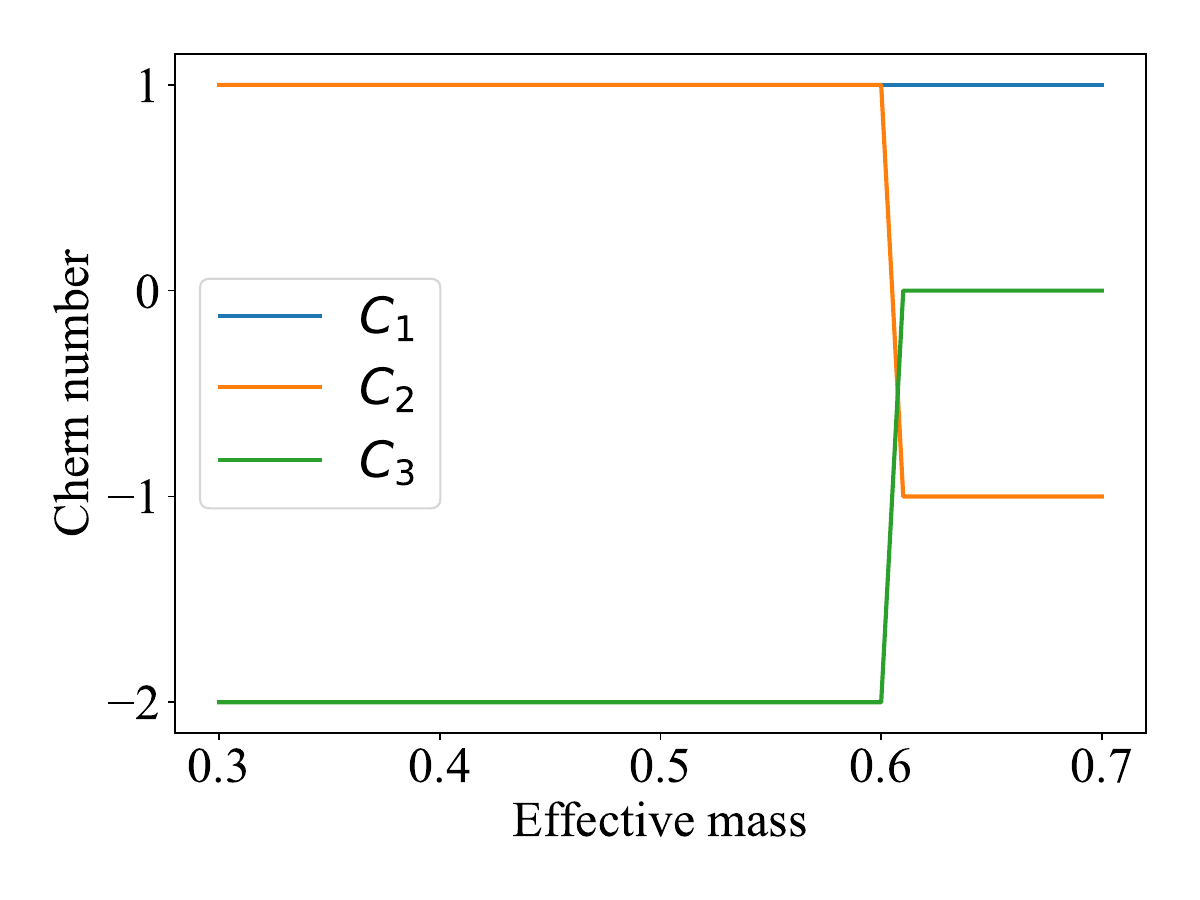}
	\caption{Phase diagram of Chen number as a function of effective mass. At a twist angle of 1.8\degree, the variation of Chern numbers for the top three bands in twisted MoTe$_2$ as the effective mass changes from 0.3 to 0.7, with other parameters taken from the reference~\cite{PhysRevLett.122.086402}.}
	\label{ChernMass}
\end{figure*}

Different from displacement field, we can see from Eq. \ref{eq:effmass} that pressure affects not only intralayer moir\'e potential $V$ and interlayer coupling $\omega$, but also the effective mass $m^*$. The evolution of bands' Chern numbers can be described in terms of a slowly varying layer-polarization texture $\Delta(\boldsymbol{r})$~\cite{zhangPolarizationdrivenBandTopology2024}:
\begin{equation}\label{eq:Delta}
	\Delta{(\boldsymbol{r})=(\mathrm{Re}t,-\mathrm{Im}t,\frac{V_b-V_t}{2}+V_z)}.
\end{equation}
Within the physical picture of layer pseudospin, a displacement field only affects the Vz term in its z-component, while pressure modifies all three components of \ref{eq:Delta}. As shown in Fig. \ref{ChernMass}, the Chern numbers of the top three energy bands will correspondingly vary when the effective mass changes.

\section{Quantum Metric Distribution}

\begin{figure*}[h]
	\includegraphics[width=1\textwidth]{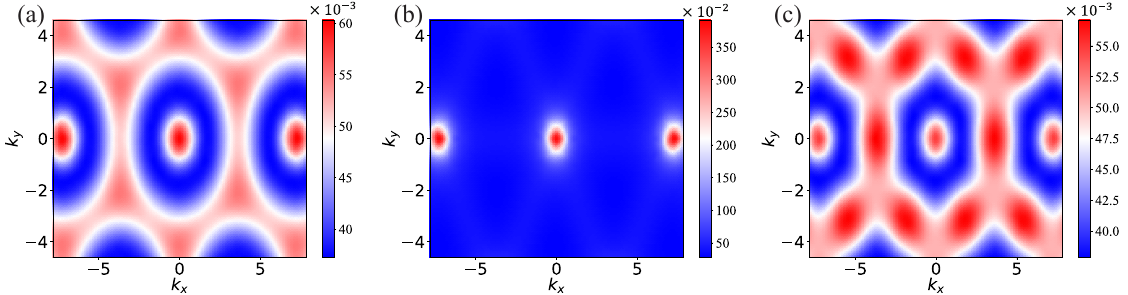}
	\caption{The tr(g($\boldsymbol{k}$)) distribution with (a) P=0 GPa, (b) P=1.8 GPa, (c) P=3.0 GPa. The twist angle is 1.8\degree.}
	\label{QM}
\end{figure*}

Figure \ref{QM} presents the distribution of quantum metric trace tr($g(\boldsymbol{k})$)=$g_{xx}(\boldsymbol{k})+g_{yy}(\boldsymbol{k})$, which shows a similar distribution to the Berry curvature in the main text. Integrating tr($g(\boldsymbol{k})$) over the mBZ, we find that for  Figs. \ref{QM}(a)-(c) $(1/2\pi)\int_{mBZ}d^2\boldsymbol{k}\ \mathrm{tr}(g(\boldsymbol{k}))$ are respectively 1.089, 1.211, 1.122. Among them, the result for P=1.8 GPa deviates the most from the LLL condition $(1/2\pi)\int_{mBZ}d^2\boldsymbol{k}\ \mathrm{tr}(g(\boldsymbol{k}))$=1.

\section{Projected Coulomb Interaction\label{coulobproject}}
The interaction Hamiltonian written in real space is given by
\begin{equation}
	H=-\int d^{2}r\sum_{\sigma,l,l'}c_{\sigma,l,\boldsymbol{r}}^{\dag}[H_0]_{ll'}c_{\sigma,l',\boldsymbol{r}}\\
	+\frac{1}{2}\int d^{2}rd^{2}r'\sum_{\sigma,\sigma',l,l'}c_{\sigma,l,\boldsymbol{r}}^{\dag}c_{\sigma',l',\boldsymbol{r'}}^{\dag}V(\boldsymbol{r}-\boldsymbol{r'})c_{\sigma',l',\boldsymbol{r'}}c_{\sigma,l,\boldsymbol{r}}.
\end{equation}
Considering we are in the hole-doping case, we should add a minus sign in front of the $H_0$. After Fourier transform, $H$ in the momentum space can be expressed as
\begin{equation}
	\tilde{H}=-\sum_{\boldsymbol{k}\in mBZ,\sigma}\epsilon_{\boldsymbol{k},\sigma}\gamma_{\boldsymbol{k},\sigma}^{\dag}\gamma_{\boldsymbol{k},\sigma}+\frac{1}{2A}\sum_{\{\boldsymbol{k_{i}}\},\sigma,\sigma'}\sum_{\boldsymbol{q}}V^{{\boldsymbol{k_{1}k_{3}},\sigma}}_{\boldsymbol{k_{2}k_{4}},\sigma'}(\boldsymbol{q})\gamma_{\boldsymbol{k_{1}},\sigma}^{\dag}\gamma_{\boldsymbol{k_{2}},\sigma'}^{\dag}\gamma_{\boldsymbol{k_{4}},\sigma'}\gamma_{\boldsymbol{k_{3}},\sigma},
\end{equation}
where we use a new set of operators $\gamma_{\boldsymbol{k},\sigma}^{\dag}=\sum_{\boldsymbol{G},l}u_{\boldsymbol{G},l,\sigma}(\boldsymbol{k})c_{\boldsymbol{k}+\boldsymbol{G},l,\sigma}^{\dag}$ to project the interactions to the top band, and $|\boldsymbol{k},\sigma\rangle=\gamma_{\boldsymbol{k},\sigma}^{\dag}|0\rangle=\sum_{\boldsymbol{G},l}u_{\boldsymbol{G},l,\sigma}(\boldsymbol{k})c_{\boldsymbol{k}+\boldsymbol{G},l,\sigma}^{\dag}|0\rangle$. Under this new basis, the kinetic part of Hamitonian is diagonalized, and $u_{\boldsymbol{G},l,\sigma}$ is the top valence band component of the eigenvector after diagonalizing the single-particle Hamiltonian. The interaction part of the Hamiltonian is
\begin{equation}
	H_{int}=\frac{1}{2A}\sum_{\boldsymbol{q}}V(\boldsymbol{q}):\rho(\boldsymbol{q})\rho(-\boldsymbol{q}):,
\end{equation}
in which the "$: :$" denotes the normal ordering of the creation and annihilation operators. The matrix elements of the density operator $\rho(\boldsymbol{q})$ are given as follows,
\begin{equation}
	\begin{aligned}
		\langle\boldsymbol{k}\sigma|\rho(\boldsymbol{q})|\boldsymbol{k'}\sigma'\rangle&=\langle\boldsymbol{k}\sigma|e^{-i\boldsymbol{q}\cdot\boldsymbol{r}}|\boldsymbol{k'}\sigma'\rangle\\&=\sum_{\boldsymbol{G_{k}},l}\sum_{\boldsymbol{G}_{\boldsymbol{k'}},l'}u_{\boldsymbol{G_{k}},l,\sigma}^{*}(\boldsymbol{k})u_{\boldsymbol{G}_{\boldsymbol{k'}},l',\sigma'}(\boldsymbol{k'})\delta_{\boldsymbol{k},[\boldsymbol{k'}+\boldsymbol{q}]}\delta_{\boldsymbol{G_{k}},\boldsymbol{G_{k'}}+\boldsymbol{G_{k'+q}}}\delta_{\sigma,\sigma'}\delta_{l,l'}\\&=\sum_{\boldsymbol{G_{k'}},l}u_{\boldsymbol{G_{k'}}+\boldsymbol{G_{k'+q}},l,\sigma}^{*}(\boldsymbol{k})u_{\boldsymbol{G_{k'}},l,\sigma}(\boldsymbol{k'}).
	\end{aligned}
\end{equation}
Here, we use the notation $[\boldsymbol{k'}+\boldsymbol{q}]$ to fold the vector $\boldsymbol{k'}+\boldsymbol{q}$ back to the first mBZ, and the relation $\boldsymbol{k}=[\boldsymbol{k'}+\boldsymbol{q}]$ must be satisfied. The folding process has the relation $\boldsymbol{k'}+\boldsymbol{q}=[\boldsymbol{k'}+\boldsymbol{q}]+\boldsymbol{G_{k'+q}}$, where $\boldsymbol{G_{k'+q}}$ is the corresponding reciprocal moir\'e vector. We set
\begin{equation}
	F(\boldsymbol{k},\boldsymbol{k'},\boldsymbol{G_{k}},\sigma)=\sum_{\boldsymbol{G_{k'}},l}u_{\boldsymbol{G_{k'}}+\boldsymbol{G_{k}},l,\sigma}^{*}(\boldsymbol{k})u_{\boldsymbol{G_{k'}},l,\sigma}(\boldsymbol{k}'),
\end{equation}
which is the form factor. The matrix elements of the density operator can be expressed as
\begin{equation}
	\rho_{\boldsymbol{k},\boldsymbol{k'},\sigma}(\boldsymbol{q})=F(\boldsymbol{k},\boldsymbol{k'},\boldsymbol{G}_{\boldsymbol{k'}+\boldsymbol{q}},\sigma).
\end{equation}

Then we can write the full expression of the matrix elements of the interaction Hamiltonian $V^{{\boldsymbol{k_{1}k_{3}},\sigma}}_{\boldsymbol{k_{2}k_{4}},\sigma'}$
\begin{equation}
	V^{{\boldsymbol{k_{1}k_{3}},\sigma}}_{\boldsymbol{k_{2}k_{4}},\sigma'}=\frac{2\pi e^{2}}{\epsilon|\boldsymbol{q}|}F(\boldsymbol{k_{1}},\boldsymbol{k_{3}},\boldsymbol{G_{k_{3}+q}},\sigma)F(\boldsymbol{k_{2}},\boldsymbol{k_{4}},\boldsymbol{G_{k_{4}-q}},\sigma').
\end{equation}

Here, we have used the momentum conservation relations $\boldsymbol{k_{1}}=[\boldsymbol{k_{3}}+\boldsymbol{q}]$ and $\boldsymbol{k_{2}}=[\boldsymbol{k_{4}}-\boldsymbol{q}]$.

\section{Cluster Shape\label{clustershape}}
\begin{figure*}[t]
	\includegraphics[width=1\textwidth]{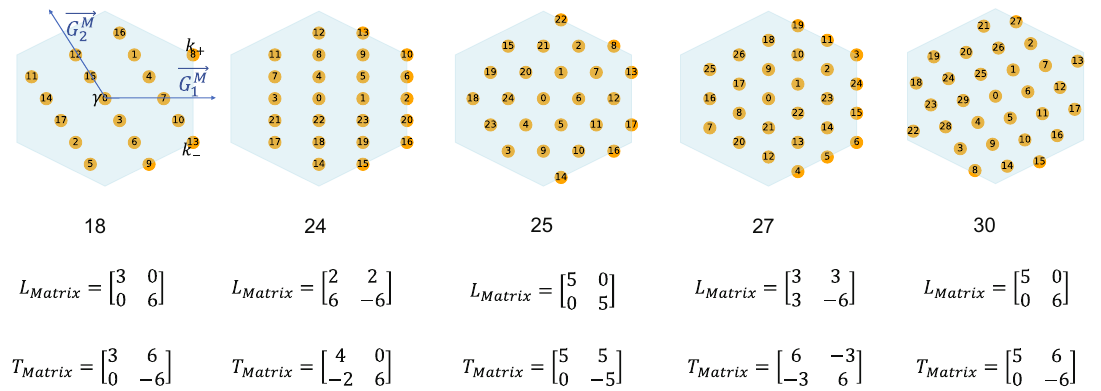}
	\caption{The different cluster shapes used in this work. $L_{Matrix}$ is used to expand the moir\'e unit cell in the real space. $T_{Matrix}$ can transform the vectors in the momentum space to the mBZ.}
	\label{cluster}
\end{figure*}

The total number of flux quanta is labeled as $N_\Phi=N_1\times N_2$, so the filling factor is $\nu=\frac{N_{e}}{N_{\Phi}}$, where $N_{i}$ is the multiple of the moir\'e lattice vector along the $\boldsymbol{a}_{i}^{M}$ and $N_e$ is the number of particles. The cluster geometries we used in this work are shown in Fig. \ref{cluster}. The basis vectors of the magnetic unit cell can be denoted as
\begin{equation}
	\begin{pmatrix}\boldsymbol{L}_1\\\boldsymbol{L}_2\end{pmatrix}=L_{Matrix}\begin{pmatrix}\boldsymbol{a}^M_1\\\boldsymbol{a}^M_2\end{pmatrix},
\end{equation} 
where $\boldsymbol{a}^M_i$ is moir\'e vector. Thus the reciprocal basis vectors can be expressed as $\boldsymbol{T}_1=2\pi*\frac{\boldsymbol{L}_{2}\times\hat{\boldsymbol{z}}}{|\boldsymbol{L}_{1}\times\boldsymbol{L}_{2}|}$ and $\boldsymbol{T}_{2}=-2\pi*\frac{\boldsymbol{L}_{1}\times\hat{\boldsymbol{z}}}{|\boldsymbol{L}_{1}\times\boldsymbol{L}_{2}|}$. The center of mass (COM) momentum of $N_e$ particles is given by $\boldsymbol{K}_\mathrm{COM}=k_{1}\boldsymbol{T}_{1}+k_{2}\boldsymbol{T}_{2}$, and can be denoted as $K_\mathrm{COM}=k_{1}+N_{1}k_{2}$ ($k_{i}=0,...,N_{i}-1$). Since $\boldsymbol{q}$ is a vector that ranges the entire selected truncated space, we need to fold $\boldsymbol{k}+\boldsymbol{q}$ back to the mBZ. Here we define a $T_{Matrix}$ which builds the relation between reciprocal vector $\boldsymbol{T}_{i}$ and the moir\'e supercell reciprocal vector $\boldsymbol{G}^M_i$
\begin{equation}
	\begin{pmatrix}\boldsymbol{G}^M_1\\\boldsymbol{G}^M_2\end{pmatrix}=T_{Matrix}\begin{pmatrix}\boldsymbol{T}_{1}\\\boldsymbol{T}_{2}\end{pmatrix},
\end{equation}
In the actual calculation, we can use the transfer matrix $T_{Matrix}$ to obtain the folded vector $[\boldsymbol{k}+\boldsymbol{q}]$ and the corresponding $\boldsymbol{G_{k+q}}$, and we have the relation $\boldsymbol{k}+\boldsymbol{q}=[\boldsymbol{k}+\boldsymbol{q}]+\boldsymbol{G_{k+q}}$. Assume that
\begin{equation}
	\boldsymbol{k}+\boldsymbol{q}=\begin{pmatrix}k_{q1} & k_{q2}\end{pmatrix}\begin{pmatrix}\boldsymbol{T}_{1}\\
		\boldsymbol{T}_{2}
	\end{pmatrix}.
\end{equation} 
$\boldsymbol{G_{k+q}}$ can be obtained by
\begin{equation}
	\boldsymbol{G_{k+q}}=\lceil\begin{pmatrix}k_{q1} & k_{q2}\end{pmatrix}T_{Matrix}^{-1}\rceil\begin{pmatrix}\boldsymbol{G}^M_1\\
		\boldsymbol{G}^M_2
	\end{pmatrix}.
\end{equation}
Here the symbol $\lceil.\rceil $ represents the greatest integer less than or equal to the argument. Then we can obtain $[\boldsymbol{k}+\boldsymbol{q}]=\boldsymbol{k}+\boldsymbol{q}-\boldsymbol{G_{k+q}}$. The respective $L_{Matrix}$ and $T_{Matrix}$ are shown in Fig. \ref{cluster}.

\section{Degenaracies of different clusters}
The degeneracy momentum counting rule for FCI is related to its particle number and cluster shape~\cite{bernevigEmergentManybodyTranslational2012}. Considering the many-body center-of-mass (COM) and relative translation symmetries on the torus, we can analyze the many-body spectrum in folded BZ of $N_{1}'\times N_{2}'$. $N_{1/2}'$ is the greatest common divisor (GCD) of $N_{1/2}$ and electron number $N_e$
\begin{equation}
	\begin{aligned}
		N_1'&=GCD(N_1,N_e),\\
		N_2'&=GCD(N_2,N_e).
	\end{aligned}
\end{equation}
We can define
\begin{equation}
	\frac{N_e}{N_1}=\frac{p_1}{q_1},\ \ \ \ \  \frac{N_e}{N_2}=\frac{p_2}{q_2}
\end{equation}

For 8 holes in $N_{\Phi_1}=4\times 6$ cluster, the folded BZ is given by $4\times 2$, the COM degeneracy mismatch is $q/(q_1q_2)=1$. Therefore the 3-fold degeneracies distribute at three different momenta in the mBZ. For 9 holes in $N_{\Phi_2}=3\times 9$, the folded BZ is still $3\times 9$ and COM mismatch is 3, so the three nearly degenerate points are located at the same momentum.

\section{Many-body Chern number and Spectral Flow\label{Many-body Chern}}
\begin{figure*}[h]
	\includegraphics[width=1\textwidth]{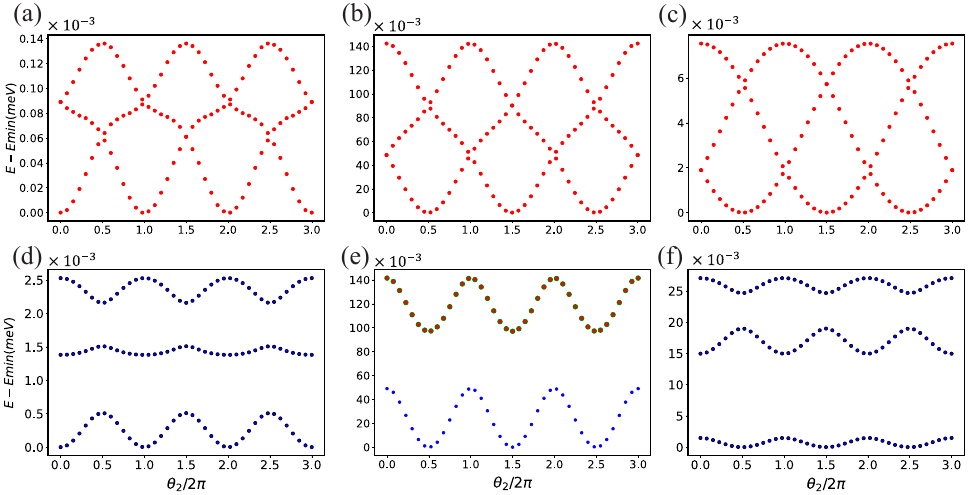}
	\caption{(a)-(c) The spectral flow of 24 sites at pressures of (a) P=0 GPa, (b) P=1.8 GPa, and (c) P=3.0 Gpa. (d)-(e) The spectral flow of 27 sites under pressure, corresponding to (a)-(c). At 1.8 GPa, the energies of $k_{+}$ and $k_{-}$ of 27 sites are degenerate. The twist angle is 1.8\degree.}
	\label{flow}
\end{figure*}

In the presence of a nonzero magnetic flux, the vector potential can be chosen to satisfy the boundary condition, 
\begin{equation}
	\boldsymbol{A_{j}}(\boldsymbol{r}_{a}+\boldsymbol{L}_{i})=\boldsymbol{A_{j}}(\boldsymbol{r}_{a})+\partial_{i}\phi_{j}(\boldsymbol{r}_{a}),
\end{equation}
where $\boldsymbol{r}_{a}$ represents the position of $a_{th}$ particle. After traversing a periodic magnetic unit cell, the wave function retains its original form up to a phase factor, as described by the transformation $\hat{T}(\boldsymbol{L}_{i})\Phi(\boldsymbol{r}_{a})\rightarrow e^{i\phi_{i}(\boldsymbol{r_{a}})}\Phi(\boldsymbol{r}_{a})$. Here, $T_{\alpha}(\boldsymbol{L}_{i})$ corresponds to the magnetic translation operator.
To compute the many-body Chern number, twisted boundary conditions must be enforced on the single-particle wave functions $\Phi(\boldsymbol{r}_{a})$
\begin{equation}
	\hat{T}(\boldsymbol{L}_{i})\Phi(\boldsymbol{r}_{a})=e^{i\theta_{i}}\Phi(\boldsymbol{r}_{a}).
\end{equation}
Thus, the applied twist phase $\theta_{i}$ is physically equivalent to introducing a magnetic flux of magnitude $\theta_{i}$.
We can redistribute the effect of the gauge transformation at the boundary evenly across all lattice sites contained within $\boldsymbol{L_{i}}$ through an appropriate transformation,
\begin{equation}
	\hat{T}(\boldsymbol{L_{i}}/N_{i})\Phi(\boldsymbol{r}_{a})=e^{i\theta_{i}/N_{i}}\Phi(\boldsymbol{r}_{a}).
\end{equation}
For each point $\boldsymbol{k}$ on the momentum-space grid, the phase is evenly distributed in each direction and is related to the momentum transition
\begin{equation}
	\delta\boldsymbol{k}_{i}=\frac{\delta\theta_{i}}{2\pi}\boldsymbol{T_{i}}.
\end{equation}
Based on this, we can compute the spectral flow during the process of increasing the magnetic flux. Figure \ref{flow} shows the spectral flow of 24 and 27 sites. For the 24-site case (Figs. \ref{flow}(a)-(c)), the spectral flows of three degenerate states along $\boldsymbol{T_2}$ exhibit level crossings at 0 GPa, 1.8 GPa, and 3.0 GPa. In contrast, as shown in Figs. \ref{flow}(d)-(f), the spectral levels for the 27-site case exhibit clear level repulsion. The absence of level crossing is attributed to the fact that the three nearly-degenerate states reside in the same momentum sector. Upon the introduction of flux, the wave functions of the three states transform identically. Without symmetry breaking or coupling between energy levels, level crossings are prohibited\cite{thoulessLevelCrossingFractional1989}. In the two cases, there is no significant difference in the flow plots at three pressures. Therefore, further analysis on the many-body Chern number is required.

In the magnetic flux parameter space defined above, we can obtain the many-body Chern number,
\begin{equation}
	\mathcal{C}_\mathrm{MB}=\frac{1}{2\pi}\int_{0}^{2\pi}d\theta_{1}\int_{0}^{2\pi}d\theta_{2}\ \mathrm{Im}(\langle\frac{\partial\Psi(\boldsymbol{\theta})}{\partial\theta_{2}}|\frac{\partial\Psi(\boldsymbol{\theta})}{\partial\theta_{1}}\rangle-c.c.).
\end{equation}
According to the definition, the partial derivative can be written as $(\Psi(\boldsymbol{\theta})-\Psi(\boldsymbol{\theta}+\boldsymbol{\delta\theta_{i}}))/|\boldsymbol{\delta\theta}|$. We can then define
\begin{equation}
	A_{i}(\boldsymbol{\theta})=\frac{\langle\Psi(\boldsymbol{\theta})|\Psi(\boldsymbol{\theta}+\boldsymbol{\delta\theta_{i}})\rangle}{|\langle\Psi(\boldsymbol{\theta})|\Psi(\boldsymbol{\theta}+\boldsymbol{\delta\theta_{i}})\rangle|}.
\end{equation}
In the discrete parameter space, we can obtain the field strength on a small plaquette~\cite{ChernNumbersDiscretized2005a},
\begin{equation}
	\Omega_{12}(\boldsymbol{\theta})=\mathrm{ln}A_{1}(\boldsymbol{\theta})A_{2}(\boldsymbol{\theta}+\boldsymbol{\delta\theta_{1}})A_{1}^{-1}(\boldsymbol{\theta}+\boldsymbol{\delta\theta_{2}})A_{2}^{-1}(\boldsymbol{\theta}).
\end{equation}
This is typically referred to as the many-body Berry curvature. Finally, the many-body Chern number can be obtained by integrating $\Omega_{12}(\boldsymbol{\theta})$,
\begin{equation}
	\mathcal{C}_\mathrm{MB}=\frac{1}{2\pi}\int_{0}^{2\pi}d\theta_{1}\int_{0}^{2\pi}d\theta_{2}\mathrm{Im}\,\Omega_{12}(\boldsymbol{\theta}).
\end{equation}

Now, let's look at the specific form of $A_{i}(\boldsymbol{\theta})$. The ground state $\Psi(\boldsymbol{\theta})$ is a linear combination of many-body orbital states: $|\Psi(\boldsymbol{\theta})\rangle=\sum_{n}U_{n}(\boldsymbol{\theta})|\psi_{n}(\boldsymbol{\theta})\rangle $. $U_{n}(\boldsymbol{\theta})$ represents the components of the eigenvectors of the many-body Hamiltonian at different COM momenta $\boldsymbol{K}$. The state $|\psi_{n}(\boldsymbol{\theta})\rangle $ can be written as the direct product of single-particle states,
\begin{equation}
	|\psi_{\boldsymbol{K}_{n}}(\boldsymbol{\theta})\rangle=\Pi_{\boldsymbol{k}}\gamma_{\boldsymbol{k}+\delta\boldsymbol{k}}^{\dag}|0\rangle=\Pi_{n}\sum_{\boldsymbol{G},\boldsymbol{l}}u_{\boldsymbol{G},\boldsymbol{l}}(\boldsymbol{k}+\boldsymbol{\delta k})c^\dag_{\boldsymbol{k}+\delta\boldsymbol{k}+\boldsymbol{G},l}|0\rangle.
\end{equation}
Here, $n$ denotes the combination set index for the particle's center-of-mass momentum $\boldsymbol{K}$, and $u_{\boldsymbol{G},\boldsymbol{l}}(\boldsymbol{k}+\boldsymbol{\delta k})$ represents the components of the eigenvector of the single-particle highest valence band. The expression for $A_{i}(\boldsymbol{\theta})$ can be expanded as,
\begin{equation}
	A_{i}(\boldsymbol{\theta})=\sum_{n}U_{n}^{*}(\boldsymbol{\theta})U_{n}(\boldsymbol{\theta}+\boldsymbol{\delta\theta_{i}})\Pi_{\boldsymbol{k}}\Pi_{\boldsymbol{k'}}\left[\delta_{\boldsymbol{k}\boldsymbol{k'}}\sum_{\boldsymbol{G},\boldsymbol{l}}u_{\boldsymbol{G},\boldsymbol{l}}^{*}(\boldsymbol{k}+\boldsymbol{\delta k})\\
	\sum_{\boldsymbol{G}',\boldsymbol{l}'}u_{\boldsymbol{G'},\boldsymbol{l'}}(\boldsymbol{k'}+\boldsymbol{\delta k'})\right].
\end{equation}
The delta function here arises from the fact that for small $\boldsymbol{\delta\theta}$, we have $\langle0|c_{\boldsymbol{k}+\boldsymbol{\delta k}}c_{\boldsymbol{k'}+\boldsymbol{\delta k'}}^{\dag}|0\rangle\approx\delta_{\boldsymbol{k}\boldsymbol{k'}}$~\cite{okamotoTopologicalFlatBands2022a}. $A_{i}(\boldsymbol{\theta})$ can be written succinctly as
\begin{equation}
	\sum_{n}U_{n}^{*}(\boldsymbol{\theta})U_{n}(\boldsymbol{\theta}+\boldsymbol{\delta\theta_{i}})\Pi_{\boldsymbol{k}}\left[\sum_{\boldsymbol{G},\boldsymbol{l}}u_{\boldsymbol{G},\boldsymbol{l}}^{*}(\boldsymbol{k}+\boldsymbol{\delta k})u_{\boldsymbol{G},\boldsymbol{l}}(\boldsymbol{k}+\boldsymbol{\delta k'})\right].
\end{equation}
The Chern number in this work is calculated on a 12$\times$12 grid for every degenerate state. When the three degenerate states are distributed to different momentum sectors, the labels 1, 2, and 3 correspond to the three states with momentum markers arranged from smallest to largest. When these states lie within the same momentum sector, the labels 1, 2, and 3 correspond to the three states with energies ordered from the lowest to the highest.

\section{Particle Entanglement Spectrum\label{entanglement}}
Particle entanglement spectrum (PES) is obtained by partitioning the system into two subsystems, A and B, containing $N_{A}$ and $N_{B}$ particles, respectively. The many-body wave function $|\Psi\rangle $ can then be expressed in its Schmidt decomposition form as

\begin{equation}
	|\Psi\rangle=\sum_{i}e^{-\frac{\xi_{i}}{2}}|\Psi_{i}^{A}\rangle\otimes|\Psi_{i}^{B}\rangle,
\end{equation}
where $\{|\Psi_{i}^{A}\rangle\}$ and $\{|\Psi_{i}^{B}\rangle\}$ form orthonormal bases in the Hilbert spaces of subsystems A and B, respectively. The total density matrix is given by $\rho= |\Psi\rangle\langle\Psi|$. For degenerate states, the total density matrix can be generalized to $\rho=\frac{1}{N_{GS}}\sum_{i}|GS_{i}\rangle\langle GS_{i}|$. Here, $N_{GS}$ is the degeneracy. Assuming that the matrix representation for $\rho$ is $P$, the reduced density matrix can be obtained as

\begin{equation}
	\rho_{A}=\mathrm{Tr}_{B}\rho=PP^{\dag}.
\end{equation}
$e^{-\xi_{i}}$ is the eigenvalue of $\rho_{A}$. $\rho $ is block-diagonal with respect to the COM momentum $\boldsymbol{K}$. Then for each sector $(\boldsymbol{K},N_{A})$, we can compute the eigenvalues of the corresponding reduced density matrix, which yield the "energy levels" of the entanglement spectrum.

\begin{figure*}[t]
	\includegraphics[width=1\textwidth]{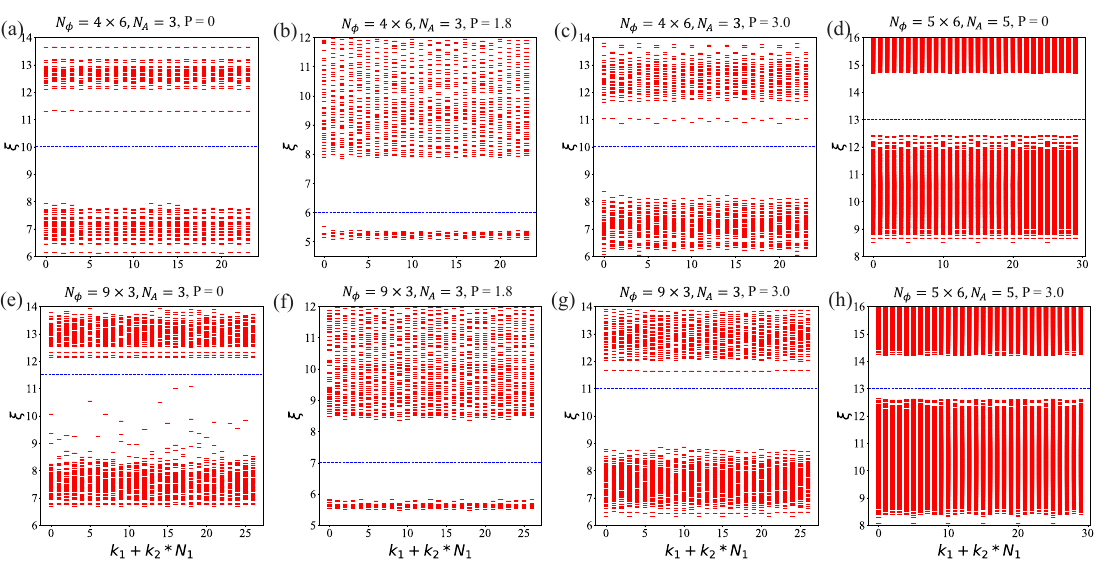}
	\caption{PES of different clusters and parameters. At P = 0 GPa, the numbers of states below the gap for $N_{\phi}=$ 24, 27 and 30 in each momentum sector are the same as P = 3.0 GPa. Here we denote the number below the dashed line in each sector as $N_{below}$. For (a)(c) $N_{\phi}$ = 24, $N_{below}$ is 46 in all the $K_{y}$ mod 3 = 0 and 45 in other sectors. For (e)(g) $N_{\phi}$ = 27, $N_{below}$ is 66 in all the $K_{y}$ mod 9 = 0 and 63 in other sectors. For (d)(h) $N_{\phi}$ = 30, $N_{below}$ is 776 in all the $K_{x}$ = 0 and 775 in other sectors. All of these results satisfy the generalized Pauli principle. At P = 1.8 GPa, the total number under the dashed lines is 168 for 24 and 252 for 27, which does not follow the FCI counting rule. }
	\label{ES}
\end{figure*}

As shown in Fig. \ref{ES}, the PES exhibits a large gap that separates the low-energy levels from the high-energy levels. In the FCI case, the number of states below the gap follows the counting rule of $\frac{1}{3}$ laughlin state~\cite{chandranBulkedgeCorrespondenceEntanglement2011}. In our results, the numbers of states below the gap for $N_{\phi} =$ 24, 27, and 30 are 1088, 1710, and 23256 at P = 0 GPa, respectively. At P = 3.0 GPa, the counting is the same as P = 0 GPa. The chosen values of $N_{A}$ are shown in Fig. \ref{ES}. These results match the quasihole counting rule for $N_{A}$ particles in the original $N_{\phi }$ lattice in the FCI~\cite{regnaultFractionalChernInsulator2011}. This implies that at P = 0 GPa and 3.0 GPa, the topmost tMoTe$_2$ valence band can host FCI. However, at P = 1.8 GPa, the numbers of states below the gap for $N_{\phi} =$ 24 and 27 are 168 and 252, which breaks the counting rule of FCI and thus exhibits the GWC.

\section{Many-Body Spectrum for other Clusters\label{other clusters}}
\begin{figure*}[t]
	\includegraphics[width=1\textwidth]{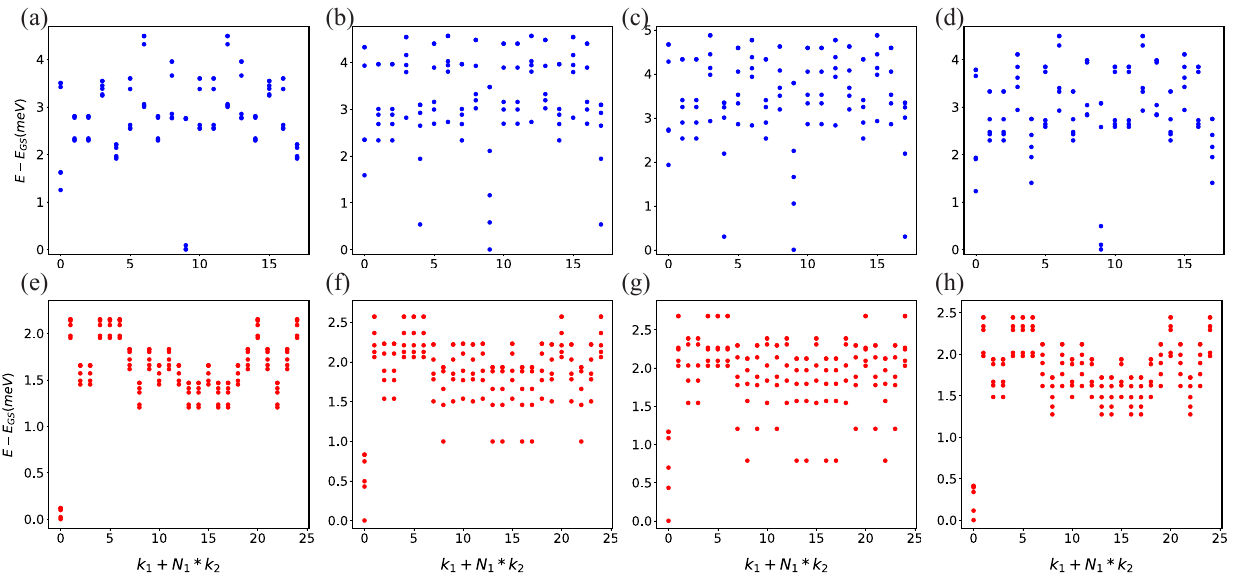}
	\caption{Additional many-body spectrums. (a)-(d)1/3 filling of 18 cluster with pressure (a) P = 0 GPa; (b) P = 1.4 GPa; (c) P = 1.8 GPa; (d) P = 3.0 GPa. (e)-(h) 2/5 filling of the 25-site cluster with the parameters corresponding to (a)-(d).}
	\label{18-25}
\end{figure*}

Here we show the many-body low energy spectrum for cluster shape of $N_{\Phi4}=18$ for $\nu=\frac{1}{3}$ and $N_{\Phi3}=25$ for $\nu=\frac{2}{5}$ of twist angle 1.8\degree, as shown in Fig. \ref{18-25}. For the 18-site cluster, the three nearly degenerate states are located in the same momentum sector at 0 GPa and 3.0 GPa, while distributed to three different momentum sectors at 1.8 GPa. For the 25-site cluster, there are five degenerate ground states at the $\gamma$ point during the pressure range P=0-1.4 GPa and 2.2-3.4 GPa. While in the range P=1.4-2.2 GPa, the system does not exhibit an energy gap separating the ground states and excited states.

\section{Many-Body Phases with Lattice Relaxation}
In our recent work, we have developed an elegant analytical framework to address the effects of relaxation in twisted systems~\cite{yu2025relaxationeffectselectronicstructure}, which effectively characterizes the influence of atomic on the electronic band properties of moir\'e structures. In this section, we will briefly introduce our methodology and key findings.

We define the atomic displacements induced by relaxation as follows: the in-plane component as $\text{\ensuremath{\boldsymbol{u_{\Vert}}^{(l)}}(\ensuremath{\boldsymbol{r}})}=u_{x}^{(l)}\hat{\boldsymbol{e_{x}}}+u_{y}^{(l)}\hat{\boldsymbol{e_{y}}}$, and the out-of-plane component as $\ensuremath{\boldsymbol{u_{\bot}}}^{(l)}(\ensuremath{\boldsymbol{r}})=u_{z}^{(l)}\hat{\boldsymbol{e_{z}}}$, where the superscript $l$ (taking values $t$ or $b$) denotes the top and bottom layers, respectively. The relaxation displacements $u$ are then determined by solving the corresponding Euler-Lagrange equations
\begin{equation}
	\frac{\partial L}{\partial u_{i}^{\pm}}-\sum_{j\in\{x,y,z\}}\partial_{j}\frac{\partial L}{\partial(\partial_{j}u_{i}^{\pm})}=0,
\end{equation}
where $u_{i}^{\pm}=u_{i}^{(t)}\pm u_{i}^{(b)}$. Then we obtain the analytical expression for atomic displacements,
\begin{equation}
	\begin{aligned}
		\boldsymbol{u}_{\Vert}^{-}(\boldsymbol{r})&=\frac{1}{\theta^{2}}\kappa_{\Vert}\sum_{j\in\{1,3,5\}}\frac{1}{||\boldsymbol{G}_{j}||^{2}}\mathrm{sin}(\boldsymbol{g}_{j}\cdot\boldsymbol{r})\boldsymbol{G}_{j},\\\boldsymbol{u}_{\bot}^{-}(\boldsymbol{r})&=\frac{1}{\theta^{2}}\kappa_{\bot}\sum_{j\in\{1,3,5\}}\frac{1}{||\boldsymbol{G}_{j}||^{2}}\mathrm{cos}(\boldsymbol{g}_{j}\cdot\boldsymbol{r})\hat{\boldsymbol{e_{z}}}.
	\end{aligned}
\end{equation}
After obtaining the relaxation displacement field, we can examine its impact on the electronic structure. It is important to note that the model parameters under strain were derived from DFT calculations on untwisted bilayer systems, during which out-of-plane relaxation was already accounted for~\cite{anfaEffectiveHamiltonianTwisted2024,PhysRevLett.122.086402}. In other words, the parameters we employed earlier inherently incorporate the effects of out-of-plane relaxation. However, this approach cannot be extended to in-plane relaxation occurring along periodic directions, and in twisted TMD systems, the influence of in-plane relaxation is significantly more substantial than that of out-of-plane relaxation. In the following, we briefly introduce the analytical method we have developed to incorporate in-plane relaxation into the electronic structure of twisted systems.

The in-plane relaxation displacement field $\boldsymbol{u}_{\Vert}(\boldsymbol{r})$, in twisted systems can be analogized as a gauge-like field. In the electronic structure, it contributes a phase factor. By expanding this phase, we map the effect of relaxation onto long-range coupling between reciprocal lattice points. Considering first-order relaxation, we derive the core relationship describing how relaxation modifies the coefficients of the continuum model---specifically, the intralayer moir\'e potential $V$ and the interlayer coupling coefficients $T$
\begin{equation}
	\begin{aligned}
		V_{X',X}^{l,(1)}(\boldsymbol{r})&=\sum_{i=1}^{6}\sum_{j=1}^{6}\nu_{X'X}^{l,i}\gamma_{ij}e^{i(\boldsymbol{g}_{i}+\boldsymbol{g}_{j})\cdot\boldsymbol{r}},\\T_{X'X}^{(1)}(\boldsymbol{r})&=\sum_{i\in\{0,2,3\}}\sum_{j=1}^{6}\omega_{X'X}\gamma_{ij}e^{i\boldsymbol{G}_{i}\cdot\boldsymbol{\tau}_{X'X}}e^{i(\boldsymbol{q}_{\eta}+\boldsymbol{g}_{i}+\boldsymbol{g}_{j})}.
	\end{aligned}
\end{equation}
Here, $X'$ and $X$ denote orbital degrees of freedom, and $\nu_{X'X}$ and $\omega_{X'X}$  are coupling coefficients. The coefficient $\gamma_{ij}=\frac{\kappa_{\Vert}}{2}\frac{1}{\theta^{2}}\frac{\boldsymbol{G}_{i}}{|\boldsymbol{G}_{i}|}\varpropto\kappa_{\Vert}$ is proportional to $\kappa_{\Vert}$, which directly determines the strength of in-plane relaxation, with $\theta$ being the twist angle. $\boldsymbol{G}_{i}$ and $\boldsymbol{g}_{i}$ represent reciprocal lattice basis vectors before and after twisting, respectively.

\begin{figure*}[t]
	\includegraphics[width=0.8\textwidth]{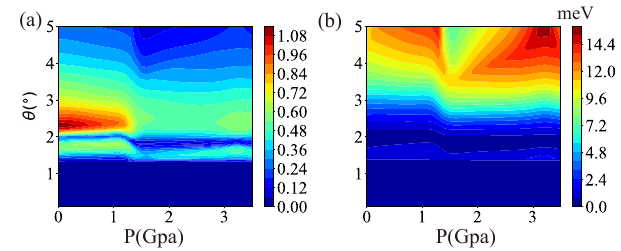}
	\caption{Figure of merit and gap between the first and second bands with lattice relaxation. (a) Figure of merit. (b) Gap between the first and second bands. Taking $\kappa_{\parallel}=2.4\times10^{-3}$, which determines the strength of in-plane relaxation.}
	\label{Relax_merit}
\end{figure*}

The magnitude of the coefficient $\kappa_{\Vert}$ is directly related to the effectiveness of relaxation. According to~\cite{PhysRevLett.124.206101,Androulidakis_2018,Advanced_Materials2019}, the value of $\kappa_{\Vert}$ typically lies in the range of 10$^{-4}$ to 10$^{-3}$. In our study, we adopted $\kappa_{\Vert}=2.4\times10^{-3}$ and computed the many-body phase diagram under relaxation. The single-particle figure of merit considering relaxation and gap between the first and second bands are shown in Figs.~\ref{Relax_merit}(a)(b). The shape of the "max merit line" is similar to the non-relaxation case, while lifting a small range of twist angles. As shown in Figs. 3(d2)-(e2), the critical twist angle for the reentrant phase transition of FCI-GWC-FCI increases from 1.8\degree in the non-relaxed case to around 2.4\degree. The analysis presented above demonstrates that pressure-tunable phase transitions between FCI and GWC persist even when lattice relaxation is taken into account.

\end{document}